\documentstyle[osa,manuscript,psfig]{revtex} 

\newcommand{\approxle}{\,{\scriptstyle\stackrel{<}{\scriptscriptstyle\sim}}\,}

\begin{document}
\titlepage

\title{Gluons in small-$x_B$ deep-inelastic scattering}

\author{Meng Ta-chung$^{1,2}$, R. Rittel$^2$, and K. Tabelow$^2$}
\address{$^1$Institute of Particle Physics, CCNU-Wuhan, 430079 Wuhan,
China\\
$^2$Institut f\"ur Theoretische Physik, FU-Berlin,
14195 Berlin, Germany\\
{\small e-mail: meng@iopp.ccnu.edu.cn; meng@physik.fu-berlin.de}}

\maketitle

\begin{abstract}
The existing data obtained in deep-inelastic scattering (DIS)
experiments in the kinematical region $x_B \approxle 10^{-2}$ and $Q^2 
\ge 5 \mbox{ GeV}^2$ are examined to gain further insight into the 
dynamics of interacting gluons. It is shown that not only the
regularities observed in diffractive scattering and those found in
large-transverse-momentum jet production, but also the striking
features of the structure function $F_2(x_B,Q^2)$ observed in normal
DIS-events in this kinematical region can be understood as direct
consequences of self-organized criticality. The results explicitly
demonstrate the usefulness of concepts and methods of Complex Sciences
in understanding the various striking features observed in high-energy
collision processes in which ``soft'' gluons play the dominating role.
\end{abstract}


Deep-inelastic lepton-nucleon scattering utilizing electron, muon,
and neutrino beams  is an extremely useful tool in studying the
structure of matter and its interactions. Since the operation of
HERA, the first electron-proton collider, the kinematical region
of such experiments  has been significantly extended. In terms of
the difference ($q$) between the initial and the final four-momenta $(k$
and $k'$) of the electron, and the four momentum $(p)$ of
the proton, the kinematical region in which data are now available is:
$0\leq Q^2\leq 9\cdot 10^4$GeV$^2$ and $10^{-5} \le x_B\leq 1$,
where $Q^2\equiv -q^2$, and $x_B\equiv -q^2/(2pq)$.

Physics of deep-inelastic lepton-nucleon scattering (DIS) in the
``small-$x_B$ region $(x_B \approxle 10^{-2}$, say)'' for sufficiently
large $Q^2$-values plays a very special role. In fact, ``Small-$x_B$
Physics'' has become one of the most active fields in Particle Physics
since the beginning of the 1990's. There are a number of reasons why
this field has become so attractive. Some of them are listed below:

($\alpha .1$) It has been reported, soon after HERA started to deliver
luminosity, by ZEUS-Collaboration \cite{ZEUSF2rise} and
H1-Collaboration \cite{H1F2rise} that, in this kinematical region, the
structure function $F_2$ rises towards smaller $x_B$, with the
strength of the rise increasing with $Q^2$ \cite{HERAreview}.  Taken
together with the fact that the valence quarks contribute very little
to $x_B \approxle  10^{-2}$, the above-mentioned experimental results
\cite{ZEUSF2rise,H1F2rise,HERAreview} imply, not only that {\em there
are many gluons in this region}, but also that {\em the interactions
of such gluons play a significant role}. Now, since in accordance with QED
and QCD, only charged particles namely the seaquark(s) and/or
antiseaquark(s) can be ``directly seen'' by the virtual photon
$\gamma^\star$, the interaction which are responsible for such
pair-creation and pair-annihilation processes cannot be neglected at
the moment when $\gamma^\star$ ``sees something'' from the gluon
system. This means, in contrast to the kinematical region $x_B\gg
10^{-2}$ where preexisting valence quarks dominate, it is {\em not
possible} to consider as first-order-approximation the struck quark or 
antiquark as a parton (which is {\em by definition} free) and to
improve the approximation by subsequently taking the relevant Feynman
diagrams into account. This means \cite{HERAreview}, theoretical
assumptions (which specify e.g. which Feynman diagrams should be
chosen, whether those leading to higher twists should be included,
etc.) are needed as input --- that is before the optimum set of
parameters in the gluon PDF's (Parton Distribution Function) can be
determined by fitting them to the experimental data
\cite{HERAreview}. 

($\alpha .2$) It is seen \cite{HERAreview}, that the above-mentioned
extension of the parton idea with the help of pQCD and empirically
determined PDF's indeed work well in reproducing the existing data
\cite{HERAreview}. Beautiful fits can be found \cite{HERAreview} for
all the measurable quantities inside as well as outside the
small-$x_B$-region ($x_B\approxle 10^{-2}$). While such fits must be
considered as excellent {\em descriptions} of the existing experimental
data \cite{HERAreview}, questions such as the following can, and should 
nevertheless be raised \cite{HERAreview}: Is this {\em the only way}
to describe Nature at the basic level of matter where
experimental information can be obtained by performing high-energy
collision processes at relatively large momentum transfer (the
so-called ``hard'' or ``semi-hard'' processes)? With such a large
number of adjustable parameters in the theory, {\em how much can we
learn about Nature} by comparing data with {\em fitted} curves? The reason
why such questions deserve to be discussed can be readily seen by
recalling what has happened in connection with the exciting discussions
about the possible quark-substructure a few years ago: The excitement
{\em began} when CDF Collaboration observed
\cite{CDFdata} a deviation between their data and a prediction based
on pQCD and the then best version of PDF's. The excitement {\em ended}
when a group of pQCD- and PDF-experts showed \cite{gluonPDF} that {\em
perfect agreement} with the data can be achieved {\em by readjusting}
some PDF-parameters concerning the gluons. Having seen this example,
we can certainly understand why more and more people are now ready to
adopt the view-point mentioned by Cooper-Sarkar, Devenish and de Roeck
in their review article \cite{HERAreview}: ``Finally there is always
the nagging doubt that the freedom to choose a fairly arbitrary
function of $x$ for the input parton distributions may be hiding the
real breakdown of the standard description.'' It is perhaps worthwhile
to compare the situation here with a well-known example in a different 
branch of
science: While everybody agrees that earthquake is catastrophic,
people may debate on ``What is more useful in {\em understanding}
Nature? Is it more useful to parameterize everthing we know about
earthquakes, insert it as input, and make use of the largest and
fastest computer to ``predict'' when, where, and how the next
earthquake comes? Or, is it more useful to discover something like the
Gutenberg-Richter law \cite{grl} although it cannot be directly used
to make ``predictions''?

($\beta$) Events with large rapidity gaps (LRG) have been observed in
this kinematical region \cite{LRGdiscovery}. This observation shows
that {\em inelastic diffractive scattering} can take place not only in
hadron-hadron collisions, but also in deep-inelastic scattering
processes. This observation gives rise to a number of questions: Are
the mechanisms which lead to such diffractive scattering processes in
different reactions related to one another? In particular, if it is
indeed ``the exchange of colorless objects'' which is responsible for
diffractive scattering in hadron-hadron collisions, is it {\em the
same kind of ``colorless objects''} which is also responsible for
diffractive scattering in deep-inelastic lepton-hadron scattering
processes \cite{LRGdiscovery,HERAreview}?  What {\em are} such
``colorless objects''? Can the existence of such objects be understood
in terms of QCD?  What is the relationship between such ``colorless
objects'' and the interacting soft gluons mentioned in ($\alpha$)?

($\gamma$) In normal DIS-events, the observed $x_B$- and $Q^2$-dependence
of $F_2$ mentioned in ($\alpha$) and the relationship between these
two properties can be well-described {\em quantitatively} by a simple
empirical formula proposed by Haidt \cite{haidt}
\begin{equation}
\label{eq1}
F_2(x_B,Q^2) = a+m \log~{Q^2\over Q_0^2} \log {x_{B0}\over x_B},
\end{equation}
where the values of the constants are $a=0.074$, $m=0.364$,
$x_{B0}=0.074$, $Q_0^2 =0.5$ GeV$^2$. As can be readily seen
\cite{haidt,thesis}, this empirical rule of Haidt \cite{haidt} indeed gives
a very good description of the data \cite{HERAreview} in the region
$x_B\approxle  10^{-2}$, for sufficiently large values of $Q^2$ ($Q^2 \ge 5$
GeV$^2$, say)! Is this empirical finding merely a parameterization
where its simplicity is nothing else but a happy coincidence? Should
this remarkably simple empirical fact about $F_2(x_B,Q^2)$ be ignored? 
Or, is it worthwhile to ask questions such as the following: Can this
empirical formula be understood in terms of QCD? What is the
relationship between this formula and the interacting soft gluons
which dominate the small-$x_B$ region?  What is the relationship
between this formula and the fact that LRG events exist in the
small-$x_B$ region?

The questions mentioned in ($\alpha$), ($\beta$), and ($\gamma$) are
closely related to one another --- so close that they should have been
discussed at the same time. But, for reasons which will be clear later
on (see iv, v, and vi below), our research began with the facts and
questions associated with those discussed in ($\beta$): 

In a recent Letter \cite{letter} and a subsequent longer paper
\cite{longSOC}, we proposed that the ``colorless objects'' which
manifest themselves in LRG events are color-singlet gluonic
BTW-avalanches due to self-organized criticality (SOC)
\cite{BTWoriginal,BTWcontinue}, and that optical-geometrical concepts
and methods are useful in examining the space-time properties of such
objects. The theoretical arguments and experimental facts which
support the proposed picture can be summarized as follows:

(i) The characteristic properties of the gluons --- especially the
local gluon-gluon coupling prescribed by the QCD Lagrangian, the
confinement, and the non-conservation of gluon numbers --- strongly
suggest that systems of interacting soft gluons are {\it open,
dynamical, complex } systems with {\it many degrees of freedom}, and
thus in such systems colorless and colored gluon-clusters in form of
BTW-avalanches (see below) can be formed.

(ii) It has been observed by Bak, Tang, and Wiesenfeld (BTW) 
\cite{BTWoriginal} that a wide class of {\em open, dynamical, complex
systems, far from equilibrium} evolve into self-organized critical
states, and local perturbations of such critical states may propagate
like avalanches caused by domino effects over all length scales. Such
a long-range correlation effect eventually terminates after a 
time-interval $T$, having reached a final amount of dissipative
energy, and having effected a total spatial extension $S$. The
quantity $S$ is called by BTW \cite{BTWoriginal} the ``size'', and the
quantity $T$ the ``lifetime'' of the ``avalanche'' and/or the
``cluster''. It is observed \cite{BTWoriginal,BTWcontinue} that there
are many such open dynamical complex systems in the macroscopic world,
and that the distributions $D_S$ of $S$ and the distribution $D_T$ of
$T$ of such BTW avalanches/clusters obey power laws: $D_S(S)\propto
S^{-\mu}$ and $D_T(T) \propto T^{-\nu}$, where $\mu$ and $\nu$ are
positive real constants. Such characteristic behaviors are known
\cite{BTWoriginal,BTWcontinue} as the ``fingerprints of SOC''.

(iii) In order to see whether SOC and thus BTW-avalanches can also
exist in microscopic systems at the level of quarks and gluons, a
systematic analysis \cite{longSOC} of the data \cite{HERAreview} for
diffractive deep-inelastic scattering (DIS) has been performed. The
results of the analysis can be summarized as follows: The
``SOC-fingerprints'' indeed exist in diffractive DIS in the
small-$x_B$ region ($x_B \approxle  10^{-2}$) where interacting soft gluons
play the dominating role. The size and the lifetime distributions of
the BTW-avalanches observed in such processes indeed show power-law
behaviors and the exponents are approximately $-2$. That is
\begin{eqnarray}
\label{eq2}
 &&D_S (S) \propto S^{-\mu} ,~~ \mu\approx 2\mbox{\, ,}\\
\label{eq3}
 &&D_T(T) \propto T^{-\nu} ,~~ \nu \approx 2.  
\end{eqnarray} 
which is true in all Lorentz frames. The observed power-law behavior
implies in particular that colorless gluon clusters are spatiotemporal
complexities which have {\it neither} a typical size, {\it nor} a
typical lifetime, {\em nor} a typical static structure.

(iv) The usefulness of the proposed SOC picture has been demonstrated
in [\ref{letter}] and in [\ref{longSOC}]. It is shown in particular
that simple analytical formulae for the differential cross-sections
$d\sigma/dt$ and $d^2\sigma/dt d(M_x^2/s)$ can be derived for
inelastic diffractive scattering, not only for small-$x_B$ DIS and for
photo-production but also for proton-proton and antiproton-proton
collisions. It has been pointed out that {\em color-singlet} gluon
clusters ($c_0^\star$) can be readily examined \cite{letter,longSOC}
experimentally in inelastic diffractive scattering processes, because
the interactions between the struck $c_0^\star$ and any other {\em
color-singlets} are of Van der Waal's type which are much weaker than
color forces at distances of hadron-radius. Thus not very much
momentum need to be transfered to a $c_0^\star$ by the projectile
(which can be a $\gamma^\star$, a $\gamma$, a $p$ or a $\bar{p}$) in
order to ``knock it out of the mother proton''. For this reason, by
considering inelastic diffractive scattering \cite{letter,longSOC}, we
were able to check the existence and the properties of the {\em
color-singlet} gluon-clusters. 

(v) After having seen \cite{letter,longSOC} the existence of
SOC-fingerprints, and having demonstrated \cite{letter,longSOC} the
usefulness of such concepts and methods in Particle Physics by
confronting them with the large body of experimental data
\cite{diffractiondata} on inelastic diffractive scattering where the
color-singlet and only the color-singlet and only the color-singlet
gluon-clusters can be calculated, we discussed as the next step, the
question: ``What can we say about those gluon-clusters which are {\em
not} color-singlets?'' Here, we have to recall that, due to the
(experimentally observed) SU(3) color-symmetry, most of such
gluon-clusters are expected to carry color quantum numbers; that is,
they are color-multiplets (instead of singlets) which will hereafter
be denoted by $c^\star$'s.  On the other hand, in accordance with the
(experimentally confirmed) characteristic features of the BTW
theory \cite{BTWoriginal,BTWcontinue}, {\em the existence of SOC
fingerprints in gluon systems cannot, and should not, depend on the
dynamical details of their interactions --- in particular not on the
intrinsic quantum numbers they exchange during the formation
process}. This means in particular, that the size ($S$) distribution
$D_S(S)$ and the lifetime ($T$) distribution $D_T(T)$ of the colored
gluon clusters (the $c^\star$'s) should not only exhibit power-law
behavior, but also have the same power as that found for color-singlet
counterparts in diffractive scattering processes
\cite{letter,longSOC}! Is the existence of SOC in open dynamical
complex systems {\em indeed so general}, and its characteristic
features {\em indeed so universal}?  Can such expectations be {\em
checked experimentally}?

(vi) The questions raised in item (v) has been discussed in a more
recent Letter \cite{jetSOC} where in particular the following has been
pointed out: Similar to individual quarks ($q$'s) or antiquarks
($\bar{q}$'s), colored gluon-clusters ($c^\star$'s) can also be
``knocked out'' of the mother proton $p$ by a projectile, {\em
provided that the corresponding transfer of momenta is large enough}.
However, in contrast to the knocked-out $q$'s or $\bar{q}$'s (in usual
DIS-events), the knocked-out $c^\star$'s may or may not have ``color
lines'' connected to the remnant of the proton --- depending on the
color-quantum number carried by the final state of the struck
$c^\star$, after it absorbs the projectile which in proton-proton
($pp$) or proton-antiproton ($p\bar{p}$) collisions can be a quark or
an antiquark or a gluon. Such knocked-out $c^\star$'s manifest
themselves in form of hadronic jets. In fact, it has been explicitly
shown in \cite{jetSOC} that the observed high-$E_T$-jets in $\bar{p}p$
collisions can be described by the proposed SOC-picture. The same idea
and the same method can be applied to lepton-induced DIS in which the
large momentum transfer is delivered by the virtual photon
$\gamma^\star$. The obtained results will be shown elsewhere.

We note in deriving the results presented in \cite{jetSOC} the
following properties of the colored gluon clusters in form of
BTW-avalanches have been explicitly taken into account:

First, because of the universality and the robustness of SOC, the
formation processes and the properties of the BTW-avalanches, in
particular the SOC-fingerprints are expected to be {\em independent}
of the intrinsic quantum number they carry. Having these and the
following arguments in mind, we conclude that the size- and the
lifetime-distributions of the colored gluonic BTW-avalanches are
expected to be {\em the same} as those for colorless ones which have
already been experimentally examined in diffractive scattering. It is
useful to note that, inside a BTW-avalanche every constituent is in
general interacting with more than one of its neighboring
constituents, in fact, everyone of them is interacting directly or
indirectly with everyone else through color forces. Hence, interaction
between a BTW-avalanche and an incident space-like photon
$\gamma^\star$ (or a hadron $h$) is weaker than the (average)
interaction between the constituents of the BTW-avalanche. This means,
in a collision process between $\gamma^\star$ (or $h$) and a
BTW-avalanche the latter acts as entire object --- {\em independent}
of the fact whether the interaction between the struck BTW-avalanche
$c^\star$ and its neighbors is of Van der Waal's type. In other words,
{\em the question whether a BTW-avalanche (either colorless or
colored) can be knocked-out depends solely on the momentum-transfer it
obtains in the collision process.}

Second, since gluonic BTW-avalanches are spatiotemporal complexities
which have neither a typical size, nor a typical lifetime, nor a
static structure, the scattering process between the virtual photon
$\gamma^\star$ and the struck avalanche $c^\star$ is {\em not} a
process in which ``the electromagnetic structure of that avalanche''
is being probed! The $\gamma^\star$ interacts within the allowed time
interval with more than one of the charged constituents of $c^\star$,
where the dynamical details about the individual subprocesses are
rather unimportant. Roughly speaking, {\em the role played by
$\gamma^\star$ is simply to deliver a sufficient amount of
momentum-transfer to the gluonic avalanche $c^\star$ such that the
entire $c^\star$ can be knocked-out of the mother proton.}

Let us now turn our attention to the majority of normal DIS-events, in
which the virtual photon $\gamma^\star$ encounters a gluonic avalanche
$c^\star$. We note, based on the facts mentioned above, it is
important to know the probability for a $\gamma^\star$ to meet such a
$c^\star$ of size $S$ and lifetime $T$; and this can be immediately
written down as the product of $D_S(S)D_T(T)$ and $ST$ which is the
space-time volume of the spatiotemporal complexity $c^\star$. Since
$F_2(x_B\approxle 10^{-2}, Q^2\ge 5 \mbox{ GeV}^2)$ is in fact nothing
else but the total probability for the above-mentioned interaction in
the given kinematical range to take place, we need to collect all
those terms which contribute to the total probability.  In doing so,
we are led to the conclusion:
\begin{eqnarray}
\label{eq4}
F_2 (x_B,Q^2)&&\,=\nonumber \\
            N&&\int_{S_{min}}^ {S_{max}} dS
               \int_{T_{min}}^{T_{max}} dT D_S(S) D_T(T)S T\mbox{\,,}
\end{eqnarray}
where $D_S(S)$ and $D_T(T)$ are given by Eqs.(\ref{eq2}) and
(\ref{eq3}) respectively, $N$ is a normalization constant, and the
integration limits are functions of $x_B$, $Q^2$, and $P\equiv |\vec{P}|$:
\begin{eqnarray} 
\label{eq5}
S_{max} &=& x_{BO} P, \rule{12mm}{0cm} S_{min} =x_{B} P,\\
\label{eq6}
T_{max} &=& \frac{4 P}{Q^2_{0}} \frac{x_B}{1-x_B},
\rule{5mm}{0cm} T_{min} =
\frac{4 P}{Q^2} \frac{x_B}{1-x_B}.
\end{eqnarray}
The yet undetermined constants $x_{B0}$ and $Q^2_0$ can be
estimated theoretically (see below).  

The facts and arguments which lead us to Eqs.(\ref{eq5}) and
(\ref{eq6}) are the following: Since whatever the charged objects hit
by $\gamma^\star(x_B,Q^2)$ may be, they are part of one of the
BTW-avalanches which dominate the small-$x_B$ region.  $S_{max}$ and
$S_{min}$ is proportional to the maximum and the minimum amount of
dissipative energy this particular BTW-avalanche can carry.  That is,
they are given by Eq.(\ref{eq5}), together with $x_{B0}
\stackrel{\scriptstyle>}{\scriptstyle\sim} 10^{-2}$ [\ref{remark1}].
From Eqs.(\ref{eq4}) and (\ref{eq5}) we see, as expected, that the
largest contribution for $F_2(x_B,Q^2)$ comes from the avalanches with
the smallest size.  Next, we consider the interaction-time
$\tau_{int}$ of such a collision event in a proper (e.g. c.m.)
Lorentz-frame: This can be estimated with the help of the Uncertainty
Principle by calculating $q^0$ (of $q\equiv k-k'$), where the result
is \cite{letter,longSOC} $\tau_{int} \equiv 1/q^0 = 4 |\vec{P}| Q^{-2}
x_B (1-x_B)^{-1}$.  The lower limit of $T$ is determined by the
requirement that the encountered BTW-avalanche has to live long enough
to been ``seen'' by the virtual photon $\gamma^*(x_B,Q^2)$. The upper
limit $T_{max}$ is determined by the requirement that the charged
object(s) which carries (carry) $x_B$ has to be recognizable by
$\gamma^\star$ (in the sense that $\gamma^\star$ should be able to
find out whether they are part of the gluon-cluster of size $S$ and
lifetime $T$). Hence the resolution power (in the transverse
directions, $1/Q^2$) of $\gamma^\star$ has to be sufficiently large
(that is $Q^2>Q_0^2$).

\begin{figure*}
\psfig{figure=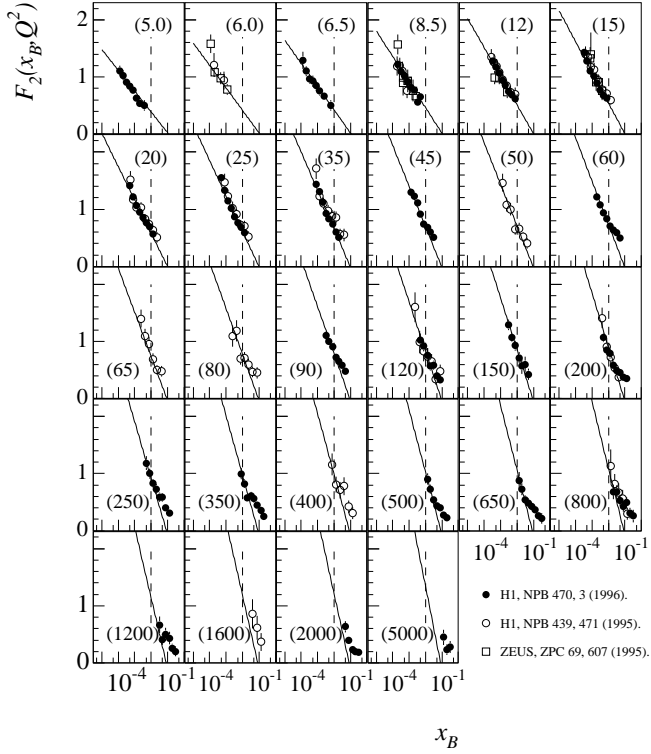}
\caption{The data for the proton structure function $F_2(x_B,Q^2)$ are 
  taken from [\ref{F2data}].  The solid lines are the calculated
  results by using the Haidt-formula [\ref{haidt}] shown in
  Eq.(\ref{eq1}). According to the theory presented in this paper, the
  formula Eq.(\ref{eq1}) should only be valid for $x_B\approxle
  10^{-2}$. This is indicated by vertical dashed lines. The $Q^2$
  values corresponding to each box are given in brackets.}
\label{fig1}
\end{figure*}

By inserting Eqs.(\ref{eq2}), (\ref{eq3}), (\ref{eq5}), and
(\ref{eq6}) into Eq.(\ref{eq4}) we obtain Eq.(\ref{eq1}) where $m$ is
related to the normalization constant $N$ in Eq.(\ref{eq4}), and $a$
can be interpreted as the averaged value of the contributions from the
valence quarks in the small-$x_B$ region \cite{remark2}.  In other
words, the empirical formula proposed by Haidt \cite{haidt}, as shown
in Eq.(\ref{eq1}), turns out to be a natural consequence of the
proposed \cite{letter,longSOC,jetSOC} SOC-picture for interacting soft
gluons which dominate small-$x_B$ deep-inelastic scattering processes.
However, while Haidt, from an experimental point of view, tried to
extend the formula (\ref{eq1}) to (if possible) all values of $x_B$,
we see here, that, from a theoretical standpoint, it is expected to be
valid only in the small-$x_B$ region. The comparison with the
experimental data is shown in Fig.  \ref{fig1}.

The authors thank their colleagues in Berlin and in Wuhan for helpful
discussions on this and other related topics. This is the revised
version of a paper with the title ``Effects of self-organized
criticality in small-$x_B$ deep-inelastic scattering''
(hep-ph/9905538). The authors thank Fu Jinghua and Liu
Lianshou for the valuable suggestions they made when this version was
written.

\begin{thebibliography}{aaa}
\bibitem{diffractiondata}
See experimental papers quoted in \cite{letter} and
\cite{longSOC}. Our apology to the authors of these papers for not
being able to repeat them here.
\bibitem{remark1}\label{remark1}
We recall that in inelastic diffractive scattering [as mentioned in
($\beta$)], the size $S$ (i.e. the dissipative energy) of the knocked-out
color-singlet gluon-cluster in form of BTW-avalanche can be directly
determined by measuring $x_P \approx M_x^2/s$ (see Refs.[\ref{letter}]
and [\ref{longSOC}] for details). Taken together with the relation
$x_B=\beta x_P$ where $0\le \beta \le 1$ (see
e.g. Refs.[\ref{LRGdiscovery}],[\ref{letter}] and
[\ref{longSOC}]), we arrive at this conclusion.
\bibitem{remark2}\label{remark2}
An estimate has been made by using the momentum-distributions for
valence quarks.  It is interesting to see that the result is of the
same order of magnitude as that given by Haidt [\ref{haidt}].
\end{thebibliography}
\end{document}